\documentclass[
reprint,
superscriptaddress,
amsmath,
amssymb,
aps,
longbibliography,
pra,
showpacs,
floatfix
]{revtex4-1}

\usepackage{graphicx}
\usepackage{color}

\usepackage{tabularx}
\usepackage{bm}
\usepackage{placeins}
\usepackage{multirow}
\usepackage{dcolumn} 
\newcolumntype{d}[1]{D{.}{.}{#1}}
\newcolumntype{L}[1]{>{\raggedright\arraybackslash}p{#1}}
\newcolumntype{C}[1]{>{\centering\arraybackslash}p{#1}}
\newcolumntype{R}[1]{>{\raggedleft\arraybackslash}p{#1}}
\usepackage{booktabs}
\usepackage{hyperref}
\definecolor{color1}{rgb}{0,0.25,0.70}

\hypersetup{colorlinks=true, 
    linkcolor={color1},
    citecolor={color1}, 
    urlcolor={color1}
}

\usepackage{comment}
\usepackage{enumitem}

\newcommand{\ang}{\ensuremath{\mathrm{\AA}}}

\begin{document}

\title{Sum-frequency excitation of coherent magnons}

\author{Dominik\ M.\ Juraschek}
\thanks{These two authors contributed equally.\\ \href{mailto:djuraschek@seas.harvard.edu}{djuraschek@seas.harvard.edu}\\
\href{mailto:prineha@seas.harvard.edu}{prineha@seas.harvard.edu}}
\author{Derek\ S.\ Wang}
\thanks{These two authors contributed equally.\\ \href{mailto:djuraschek@seas.harvard.edu}{djuraschek@seas.harvard.edu}\\
\href{mailto:prineha@seas.harvard.edu}{prineha@seas.harvard.edu}}
\author{Prineha\ Narang}
\affiliation{Harvard John A. Paulson School of Engineering and Applied Sciences, Harvard University, Cambridge, MA 02138, USA}

\date{\today}

\begin{abstract}
Coherent excitation of magnons is conventionally achieved through Raman scattering processes, in which the difference-frequency components of the driving field are resonant with the magnon energy. Here, we describe mechanisms by which the sum-frequency components of the driving field can be used to coherently excite magnons through two-particle absorption processes. We use the Landau-Lifshitz-Gilbert formalism to compare the spin-precession amplitudes that different types of impulsive stimulated and ionic Raman scattering processes and their sum-frequency counterparts induce in an antiferromagnetic model system. We show that sum-frequency mechanisms enabled by linearly polarized driving fields yield excitation efficiencies comparable or larger than established Raman techniques, while elliptical polarizations produce only weak and circularly polarizations no sum-frequency components at all. The mechanisms presented here complete the map for dynamical spin control by the means of Raman-type processes.
\end{abstract}

\maketitle



Ultrashort laser pulses are able to generate coherent quasiparticle excitations and induce macroscopic amplitudes in the normal mode coordinates, such as atomic motion for phonons and spin precession for magnons. The most prominent excitation mechanisms are Raman processes: impulsive stimulated Raman scattering, in which photons from an ultrashort laser pulse scatter with a Raman-active quasiparticle, and ionic Raman scattering, in which photons from an ultrashort pulse initially excite infrared-active phonons coherently, which then in turn scatter with a Raman-active quasiparticle. Impulsive stimulated Raman scattering is an established means of excitation for Raman-active phonons \cite{desilvestri:1985,merlin:1997} and magnons \cite{kimel:2005,Kalashnikova2007,Kalashnikova2008,Gridnev2008,Popova2012,Kalashnikova2015,Satoh2017,Tzschaschel2017}. Ionic Raman scattering in turn, predicted 50 years ago \cite{maradudin:1970,Wallis1971,Humphreys1972}, has been demonstrated and described only in recent years for Raman-active phonons \cite{forst:2011,Forst2013,subedi:2014,Mankowsky:2015} and magnons \cite{nova:2017,juraschek2:2017,Juraschek2020_3} due to the development of powerful terahertz and mid-infrared sources.

\begin{figure*}[t]
\centering
\includegraphics[scale=0.15]{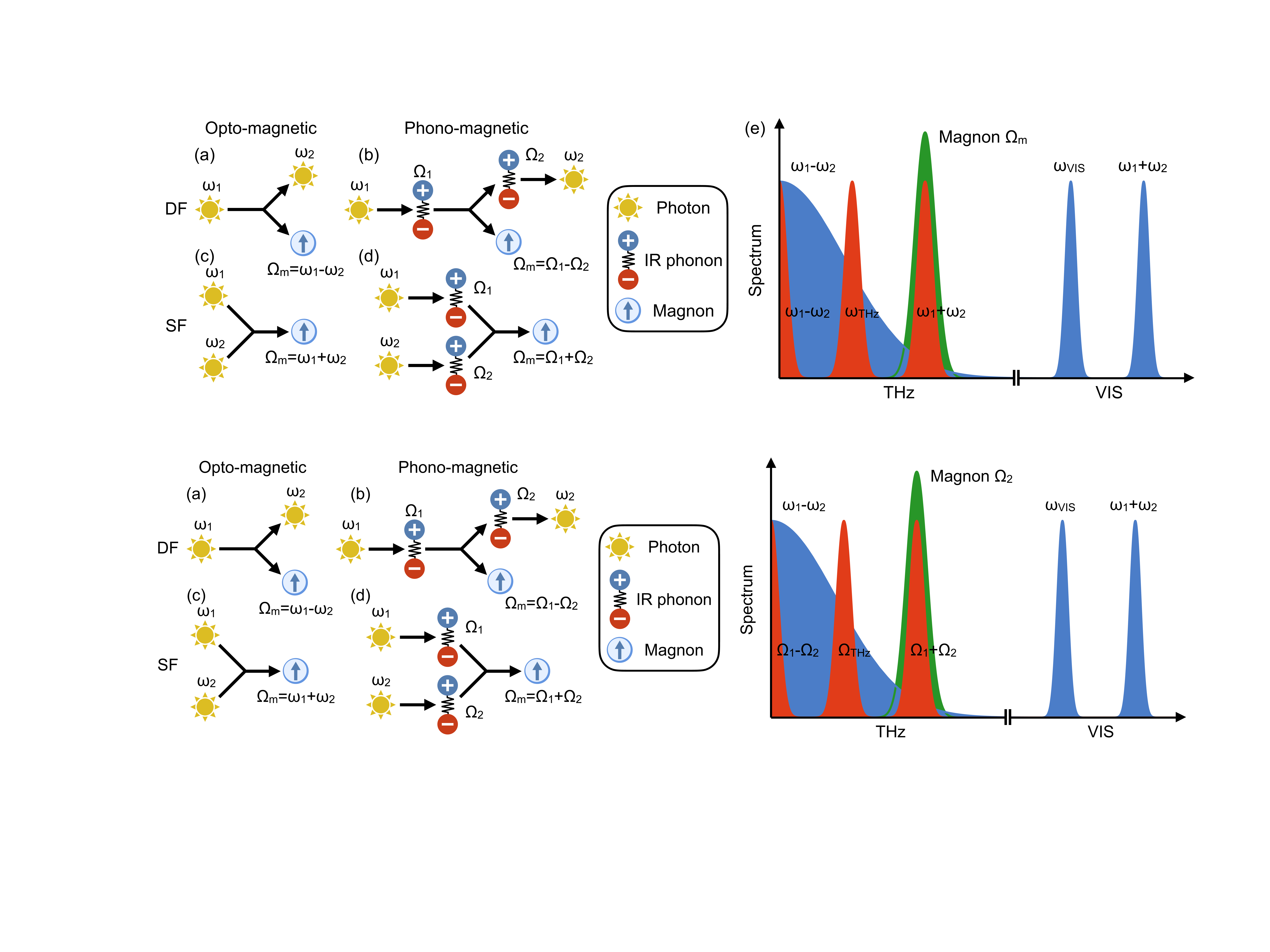}
\caption{Fundamental scattering processes of photons and coherent infrared-active phonons with magnons. (a) Impulsive stimulated Raman scattering, (b) ionic Raman scattering, (c) two-photon absorption, (d) two-phonon absorption. (e) Schematic frequency spectrum of a magnon and opto-magnetic drives in the visible and terahertz range. Shown are the frequency of the magnon, $\Omega_\mathrm{m}$ (green), the center frequencies of the visible and terahertz pulses, $\omega_\mathrm{VIS}$ (blue) and $\omega_\mathrm{THz}$ (red), as well as the difference- and sum-frequency components of the respective driving forces, $\omega_1-\omega_2$ and $\omega_1+\omega_2$. In this example, the difference-frequency components of the visible-light pulse and the sum-frequency components of the terahertz pulse overlap with the magnon mode and are therefore able to excite it.
}
\label{fig:schematic}
\end{figure*}

Phenomenologically, these Raman processes are described by an $A_\mathrm{R} A_\mathrm{D}^2$ interaction term in the free energy, where $A_\mathrm{R}$ is the amplitude of the normal mode coordinate of the Raman-active quasiparticle and $A_\mathrm{D}$ is the amplitude of the driving field. In the case of impulsive stimulated Raman scattering, $A_\mathrm{D}$ corresponds to the electric field amplitude of the laser pulse, and in the case of ionic Raman scattering, $A_\mathrm{D}$ is the vibrational amplitude of the coherent infrared-active phonon. If the driving field follows a sinusoidal shape $A_\mathrm{D}\sim\sin(\omega_0 t)$ with center frequency $\omega_0$, then the force acting on the Raman-active quasiparticle, $A_\mathrm{D}^2\sim1-2\cos(2\omega_0t)$, produces a static term and one oscillating at double the frequency. For finite linewidths, these terms consist of difference- and sum-frequency components, $\omega_1-\omega_2$ and $\omega_1+\omega_2$, respectively. In conventional Raman scattering processes, the difference-frequency components of the driving field are resonant with the frequency of the Raman-active quasiparticle, as illustrated in Figs.~\ref{fig:schematic}(a) and (b). Very recently, excitations of coherent Raman-active phonons through sum-frequency components have been achieved, in which the Raman-active phonon absorbs two photons \cite{maehrlein:2017,Juraschek2018,Knighton2019,Johnson2019} or two infrared-active phonons \cite{Juraschek2018,Melnikov2018,Kozina2018,Knighton2019,Johnson2019,Melnikov2020}. In contrast, both experimental and theoretical demonstrations are still missing for the excitation of magnons through sum-frequency processes.

In this study, we develop a phenomenological description of the sum-frequency counterparts (Figs.~\ref{fig:schematic}(c) and (d)) of impulsive stimulated and ionic Raman scattering for the coherent excitation of magnons, building on the formalism developed in Ref.~\cite{Juraschek2020_3}. The sum-frequency description of the scattering process differs conceptually from the Raman scattering processes with phonons \cite{maehrlein:2017,Juraschek2018}, because angular momentum considerations have to be taken into account. We compare the relative strengths of the difference- and sum-frequency mechanisms for the example of an anisotropic antiferromagnetic Heisenberg model, for which we evaluate the spin-precession amplitudes for different optical and phononic drives. We show that sum-frequency mechanisms enabled by linearly polarized driving fields yield excitation efficiencies comparable or larger than established Raman techniques for comparable coupling strengths, while elliptical polarization produces only weak and circularly polarization no sum-frequency components at all. In addition, the sum-frequency mechanisms benefit benefit from a higher selectivity due to the possibility of resonant tuning and due to the lower energy of the excitation.

\section{Theory of optically and phonon-induced spin dynamics}


\subsection{Landau-Lifshitz-Gilbert spin dynamics}

We begin by reviewing the phenomenological theory of spin dynamics in the Landau-Lifshitz-Gilbert formalism. The precession of a spin, $\mathbf{S}_s$ with $|\mathbf{S}_s|=1$, follows the equation of motion
\begin{equation} \label{eq:LLG}
    \frac{{\rm d}\mathbf{S}_s}{{\rm d}t}=-\frac{\gamma}{1+\Gamma^2}\left[\mathbf{S}_s \times \mathbf{B}_s^{\rm eff} - \frac{\Gamma}{|\mathbf{S}_s|} \mathbf{S}_s \times (\mathbf{S}_s \times \mathbf{B}_s^{\rm eff})\right],
\end{equation}
where $\gamma=g\mu_{\rm B}/\hbar$, $g$ is the material-dependent gyromagnetic ratio, $\mu_{\rm B}$ is the Bohr magneton, $\hbar$ is the reduced Planck's constant, and $\Gamma$ is a phenomenological damping constant. The effective magnetic field acting on the spin is given by $\mathbf{B}_s^{\rm eff}=\partial H/\partial \mathbf{S}_s$, where $H$ is the Hamiltonian of the system. Here, we use a model of an anisotropic Heisenberg antiferromagnet with two sublattices, $s\in \{1,2\}$, which interacts with light and coherent phonons,
\begin{equation} \label{eq:generalHamiltonian_maintext}
    H = H_0 + H_{\rm phot} + H_{\rm phon}.
\end{equation}
The ground state Hamiltonian
\begin{equation}\label{eq:groundstate}
H_0 = J \mathbf{S}_1 \cdot \mathbf{S}_2 + \sum\limits_{s=1}^2 \left[ D_x S_{sx}^2 + D_y S_{sy}^2 \right]
\end{equation}
consists of a term with antiferromagnetic exchange coupling, $J=6J_\mathrm{NN}>0$, where $J_\mathrm{NN}$ is the nearest-neighbor antiferromagnetic exchange coupling, and anisotropy terms along the $x$ and $y$ directions, $D_x$ and $D_y$, which align the spins along the $z$ direction, as shown schematically in Fig.~\ref{fig:experiment}. The studied model is similar to those used to describe the prototypical antiferromagnet nickel oxide (NiO) \cite{Hutchings1972,Kampfrath2011,Rezende2019}, but for demonstration purposes, we use artificial values for the material parameters here that do not reflect the properties of NiO. $H_{\rm phot}$ and $H_{\rm phon}$ describe the interactions of the spins with light and with coherent infrared-active phonons. In the ground state and without external magnetic fields, this model exhibits two magnon modes, one out-of-plane mode with energy $\hbar\Omega_{\rm IP} = 2\sqrt{J(D_x+D_y)}$, and one in-plane mode with energy $\hbar\Omega_{\rm OP} = 2\sqrt{JD_y}$ \cite{Rezende2019}. The effective magnetic field experienced by the spins in equilibrium, $\mathbf{B}_s^{\rm 0}$, is given by \cite{Kampfrath2011}
\begin{eqnarray} \label{eq:effectiveB}
    \mathbf{B}_s^{\rm 0} & = &  \frac{1}{\gamma\hbar}\partial H_0/ \partial \mathbf{S}_s \\
    & = & \frac{1}{\gamma\hbar} [J(S_{sx}\hat{x}+S_{sy}\hat{y}+S_{sz}\hat{z}) \nonumber \\
    & & +  2(D_x S_{sx}\hat{x}+D_y S_{sy}\hat{y})].
\end{eqnarray}

In the following sections, we describe the effective magnetic fields exerted on the spins by the electric-field components of a laser pulse (opto-magnetic effects) and by coherently excited infrared-active phonons (phono-magnetic effects). We hereby extend the formalism of Ref.~\cite{Juraschek2020_3} to include both difference-frequency components through impulsive stimulated and ionic Raman scattering, as as well sum-frequency components through two-photon and two-phonon absorption. The effective magnetic field picture provides an intuitive description of the underlying physics in the regime of stable magnetic order, where no photo- or phonon-induced melting is induced \cite{Rini2007,Tobey2008,Beaud_et_al:2009,Caviglia2013,Hu2016,Forst2017}. Note that for very short pulses, the diamagnetic response in the effective magnetic field picture may not be entirely accurate anymore and additional diamagnetic effects resulting from a quenching of orbital angular momentum have to be taken into account that may even dominate the response in some cases \cite{reid:2010,Mikhaylovskiy2012,Gorelov2013}. The scattering process then has to be treated in a fully quantum-mechanical formalism \cite{popova:2011,Popova2012,Berritta2016,Majedi2020}.


\subsection{Effective opto-magnetic fields}\label{ssec:optomagneticfield}

We now derive the effective opto-magnetic fields arising from the interaction of the electric-field components of a laser pulse with magnetic order. In conventional descriptions of nonlinear light-matter interactions in the field energy density approach, a time averaging is applied that eliminates fast-oscillating components \cite{Pershan1963, pershan:1966, Zon1983, BoydNonlinearOptics}. To obtain both difference- and sum-frequency components of the opto-magnetic field, we write the field energy density, $U$, of a non-absorbing and linearly polarized material without time averaging as
\begin{equation}
    U = \frac{1}{2} D_i(t) E_i(t),
\end{equation}
where $E_i$ the electric field and $D_i=\varepsilon_0 E_i + P_i$ the electric displacement field along the spatial coordinate $i$, and $\varepsilon_0$ is the vacuum permittivity. The complex spectral decomposition of the electric field reads $E_i(t) = \sum_n \tilde{E}_i (\omega_n) e^{-{\mathrm i}\omega_n t}$, where $\tilde{E}_i(\omega_n)$ is the amplitude of the electric-field component with frequency $\omega_n$, the sum over $n$ runs over all negative and positive frequencies, and $\tilde{E}_i(\omega_n)=\tilde{E}_i^*(-\omega_n)$ to ensure that $E_i(t)$ is real. We use the Einstein notation for the summation of indices. The polarization, $P_i$, up to first order in the electric field is given by $P_i(t) = \varepsilon_{0} \sum_n \chi_{ij}(\omega_n) \tilde{E}_j(\omega_n) e^{-{\mathrm i}\omega_n t}$, where $\chi_{ij}$ is the frequency-dependent linear susceptibility, for which $\chi_{ij}(\omega_n)=\chi_{ij}^*(-\omega_n)= \chi_{ji}^*(\omega_n)$. We accordingly rewrite the field energy density $U$ as
\begin{equation}\label{eq:Usplit}
    U = \underbrace{\frac{1}{2} \varepsilon_0 E_i(t) E_i(t)}_{H_\mathrm{vac}/V_{\rm c}} + \underbrace{\frac{1}{2} P_i(t) E_i(t)}_{H_{\rm phot}/V_{\rm c}},
\end{equation}
where $H_\mathrm{vac}$ is the Hamiltonian of the vacuum polarization, $H_{\rm phot}$ that of the linear polarization of the medium, and where $V_{\rm c}$ is the volume of the unit cell. 

$\chi_{ij}$ is further dependent on the magnetic order and we expand it up to second order in the ferromagnetic and antiferromagnetic vectors, $\mathbf{m}=\mathbf{S}_1 + \mathbf{S}_2$ and $\mathbf{l}=\mathbf{S}_1 - \mathbf{S}_2$, respectively, as \cite{Kalashnikova2015,Satoh2017,Tzschaschel2017,Juraschek2020_3}
\begin{eqnarray}\label{eq:dielectricexpansion}
    \chi_{ij} & = & \chi_{ij}^\mathrm{gs} + {\rm i}\alpha_{ijk} m_k + {\rm i}\alpha_{ijk}' l_k \nonumber \\
    & & + \beta_{ijko} m_k m_o +\beta_{ijko}' l_k l_o +\beta_{ijko}'' m_k l_o,
\end{eqnarray}
where $\chi_{ij}^\mathrm{gs}$ is the ground-state linear susceptibility, and $\alpha^{(}{'}{}^{)}$ and $\beta^{(}{'}{}^{,}{''}{}^{)}$ are the frequency-dependent first and second order opto-magnetic coefficients (or magnetic Raman tensors). The coupling of the laser pulse to the magnetization in first order is known as inverse Faraday effect, while the coupling in second order is known as inverse Cotton-Mouton effect \cite{kimel:2005,Kalashnikova2007,Kalashnikova2008,Gridnev2008,Popova2012,Kalashnikova2015,Satoh2017,Tzschaschel2017,Juraschek2020_3}.

\begin{figure}[t]
\centering
\includegraphics[scale=0.145]{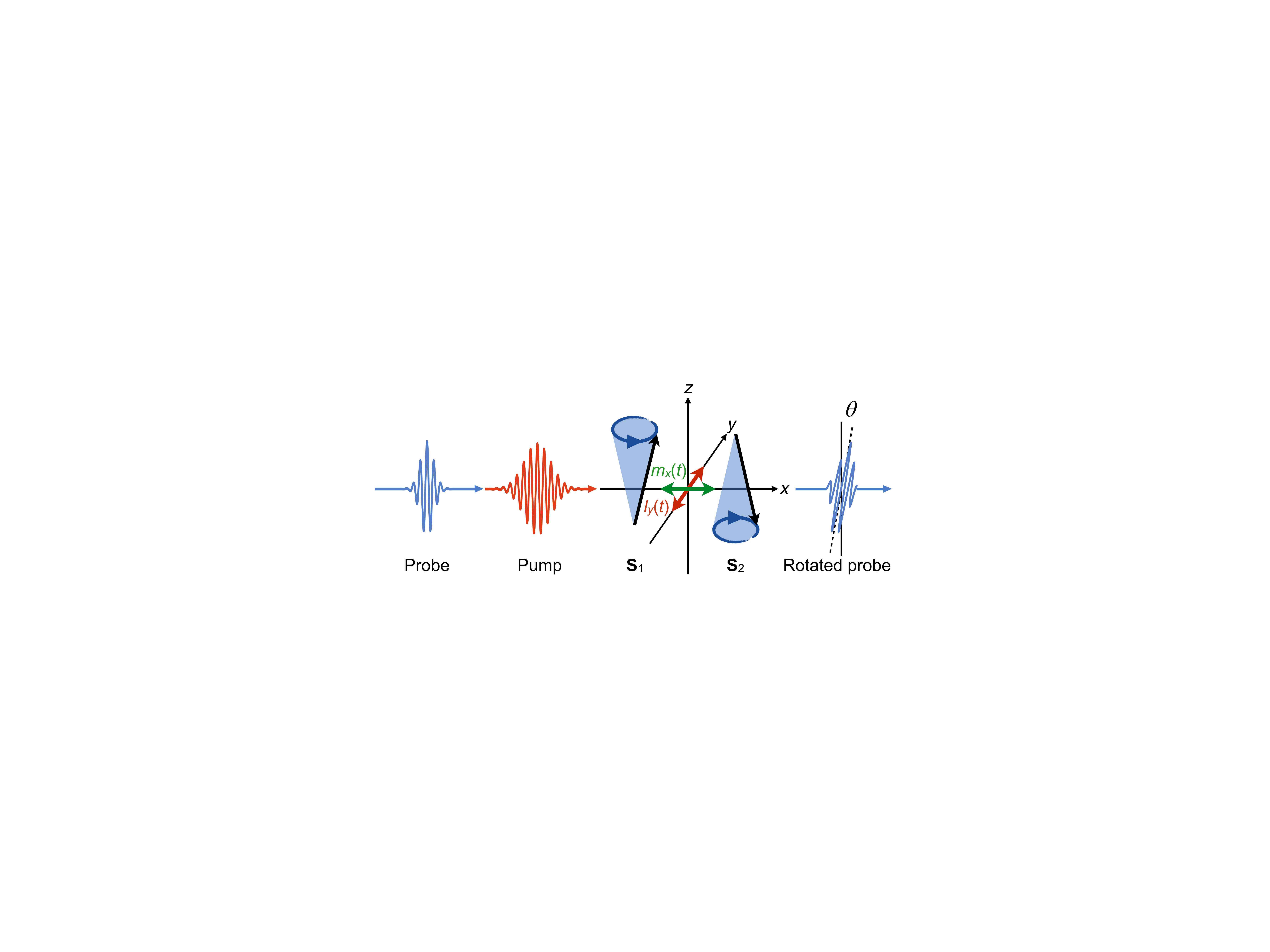}
\caption{Schematic of the pump-probe geometry. The spins of the antiferromagnetic model are aligned along the $z$-direction. The in-plane magnon mode excited by the pump pulse in this geometry (either directly, or through coherently excited phonons) distorts the spins antiferromagnetically along the $y$-direction, $l_y(t)$, as well as ferromagnetically along the $x$-direction, $m_x(t)$. The probe pulse then undergoes a Faraday rotation, $\theta(t)$, which follows the oscillation of $m_x(t)$.
}
\label{fig:experiment}
\end{figure}

We assume that our model system is irradiated with light propagating along the $x$-direction, with electric-field components lying in the $yz$-plane, as illustrated in Fig.~\ref{fig:experiment}. In this geometry, symmetry considerations prevent terms linear in $l$ \cite{Satoh2017,Tzschaschel2017}, and the $\beta_{ijko}$ term can be neglected, because it scales with $m^2$, which is small in an antiferromagnet. For the remaining coefficients, $\alpha_{zyx}=-\alpha_{yzx}$ and $\beta_{zyyz}'=\beta_{zyzy}'=\beta_{yzzy}'=\beta_{yzyz}'$. Using Eqs.~(\ref{eq:Usplit}) and (\ref{eq:dielectricexpansion}), the interaction Hamiltonian of light with magnetic order reads
\begin{eqnarray}
H_\mathrm{phot} & = & \frac{V_c\epsilon_0}{2}\Big[\chi_{yz}^\mathrm{gs}(\omega_z)-\mathrm{i} \alpha_{yzx}(\omega_z) m_x \nonumber \\ 
& + &  2\beta_{yzzy}'(\omega_z) l_y l_z \tilde{E}_z^*(\omega_z)e^{\mathrm{i}\omega_z t}+\mathrm{c.c}\Big]E_y(t) \nonumber \\
& + & \frac{V_c\epsilon_0}{2}\Big[\chi_{yz}^\mathrm{gs}(\omega_y)+\mathrm{i} \alpha_{yzx}(\omega_y) m_x \nonumber \\ 
& + &  2\beta_{yzzy}'(\omega_y) l_y l_z \tilde{E}_y^*(\omega_y)e^{\mathrm{i}\omega_y t}+\mathrm{c.c}\Big]E_z(t).
\end{eqnarray}
The effective opto-magnetic field acting on the spin $\mathbf{S}_s$ is then given by $\mathbf{B}_s^{\rm eff}=(\gamma \hbar)^{-1}\partial H_{\rm phot}/\partial \mathbf{S}_s$. 

To induce an effective opto-magnetic field through the inverse Faraday effect, light traveling in the $x$-direction needs to be elliptically or circularly polarized, and we write for the $z$ and $y$ components of the electric field
\begin{eqnarray}
    E_z(t) & = & \mathcal{E}_z(t)(e^{-{\mathrm i}\omega_z t}+e^{{\mathrm i}\omega_z t}), \label{eq:Ez} \\
    E_y(t) & = & \mathcal{E}_y(t)({\mathrm i}e^{-{\mathrm i}\omega_y t} - {\mathrm i} e^{{\mathrm i}\omega_y t}). \label{eq:Ey}
\end{eqnarray}
Here, $\mathcal{E}_i(t)=\mathcal{E}_{i0}{\rm exp}\{-t^2/[2(\tau/\sqrt{8~{\rm ln}~2})^2\}$ is a Gaussian carrier envelope, $\mathcal{E}_{i0}$ is the peak electric field along direction $i$, and $\tau$ is the full width at half maximum pulse duration. In general, $E_z(t)$ and $E_y(t)$ can have different center frequencies and their phase difference will change in time; in circularly polarized light, $\omega_z = \omega_y$ and their phase difference is $\pi/4$.

Neglecting the dependence of $\chi_{ij}$ on the shape of the carrier envelope and assuming $l_z$ is constant in time as in Ref. \cite{Tzschaschel2017}, we find that the inverse Faraday effect, mediated through the $\alpha_{zyx}$ term, generates an effective opto-magnetic field, $\mathbf{B}_s^{\rm IFE}$, according to
\begin{eqnarray}\label{eq:IFEfield}
    \mathbf{B}_s^{\rm IFE}(t) & = & \frac{\varepsilon_0 V_c}{\gamma\hbar} \mathcal{E}_z(t) \mathcal{E}_y(t) \nonumber\\
    & & \times \left[\alpha_+ \cos(\omega_-t)+\alpha_- \cos(\omega_+t)\right]\hat{x},
\end{eqnarray}
where $\mathbf{B}_1^{\rm IFE}=\mathbf{B}_2^{\rm IFE}$ and where we defined $\omega_\pm = \omega_z \pm \omega_y$ and $\alpha_\pm=\alpha_{zyx}(\omega_y)\pm\alpha_{zyx}(\omega_z)$. For circularly polarized light ($\mathcal{E}_y=\mathcal{E}_z\equiv\mathcal{E}$ and $\omega_y = \omega_z$), $\alpha_-$ vanishes and we obtain the conventional equation for the inverse Faraday effect \cite{pershan:1966,Kalashnikova2015,Tzschaschel2017},
\begin{equation}\label{eq:IFEfielddiff}
    \mathbf{B}_s^{\rm IFE}(t) = \frac{\varepsilon_0 V_c}{\gamma\hbar} \mathcal{E}^2(t) \alpha_+ \hat{x}.
\end{equation}
The inverse Faraday effect involving circularly polarized light can therefore not produce sum-frequency components. More generally it can only produce sum-frequency components if $\alpha_{zyx}$ shows a large frequency dependence in order to make $\alpha_-$ significant. Conventionally, Raman tensors are only weakly frequency dependent in spectral ranges far below the band gap \cite{Juraschek2018}, and we expect similar behavior for magnetic Raman tensors here.

For an effective opto-magnetic field induced by the inverse Cotton-Mouton effect, we assume a laser pulse travelling along the $x$-direction, which is linearly polarized in the $yz$-plane with an angle of $\phi$ with respect to the $z$ axis,
\begin{eqnarray}
    E_z(t) & = & \mathcal{E}_z(t)\cos(\phi)(e^{-{\mathrm i}\omega_z t}+e^{{\mathrm i}\omega_z t}), \label{eq:Ez_linearlyPolarized} \\
    E_y(t) & = & \mathcal{E}_y(t)\sin(\phi)(e^{-{\mathrm i}\omega_y t}+e^{{\mathrm i}\omega_y t}) \label{eq:Ey_linearlyPolarized}.
\end{eqnarray}
The effective opto-magnetic field generated by the inverse Cotton-Mouton effect, $\mathbf{B}_s^{\rm ICME}$, mediated through the $\beta_{zyyz}'$ term, then yields
\begin{eqnarray}\label{eq:ICMEfield}
    \mathbf{B}_{1/2}^{\rm ICME}(t)
    & = & \pm\frac{\varepsilon_0 V_c}{\gamma\hbar} \mathcal{E}_z(t) \mathcal{E}_y(t) \beta_+' \sin(2\phi) \nonumber\\
    & & \times \left[ \cos(\omega_+t) + \cos(\omega_-t) \right] l_z \hat{y},
\end{eqnarray}
where we defined $\beta_+' = \beta_{zyyz}'(\omega_y)+\beta_{zyyz}'(\omega_z)$. The inverse Cotton-Mouton effect is maximized when the linear polarization is oriented at 45 degrees between the $y$ and $z$ axes. Both the difference- and sum-frequency components scale identically with $\beta_+'$, in contrast to the inverse Faraday effect.


\subsection{Effective phono-magnetic fields}\label{ssec:phonomagneticfield}

Next, we derive the effective phono-magnetic fields produced by coherent infrared-active phonons analogously to the previous section. In previous work, Ref.~\cite{Juraschek2020_3}, fast-oscillating components were eliminated in the derivation of the Raman scattering mechanisms. In order to include both difference- and sum-frequency components, we write the interaction Hamiltonian, $H_{\rm phon}$, as
\begin{equation} \label{eq:phonomagneticHamiltonian}
    H_{\rm phon} =\frac{1}{2} I_i(t)Q_i(t),
\end{equation}
where $Q_i(t) = \sum_n \tilde{Q}_i (\Omega_n) e^{-{\mathrm i}\Omega_n t}$ is the phonon amplitude and $I_i(t)= \sum_n D_{ij}(\Omega_n) \tilde{Q}_j(\Omega_n) e^{-{\mathrm i}\Omega_n t}$ can be regarded as phononic displacement field. We again use the Einstein notation for summing indices, which here denote the phonon band number and not spatial coordinates. $D_{ij}=\mathbf{q}_i^\mathrm{T} D \mathbf{q}_j$ is the projected dynamical matrix, where $\mathbf{q}_i$ is the eigenvector of phonon mode $i$ and $D$ the dynamical matrix. For $i=j$, we obtain the eigenfrequency of phonon mode $i$ as $\Omega_{i}=\sqrt{D_{ii}}$. We expand $D_{ij}$ to second order in the ferro- and antiferromagnetic vectors, $\mathbf{m}$ and $\mathbf{l}$, analogously to the linear susceptibility in the previous section, yielding \cite{Juraschek2020_3}
\begin{eqnarray}\label{eq:dynamicalexpansion}
    D_{ij} & = & D_{ij}^\mathrm{gs} + {\rm i} a_{ijk} m_k + {\rm i} a'_{ijk} l_k \nonumber \\
    & & + b_{ijko} m_k m_o + b_{ijko}' l_k l_o + b_{ijko}'' m_k l_o,
\end{eqnarray}
where $D_{ij}^\mathrm{gs}$ is the projected dynamical matrix of the ground state, and $a^{(}{'}{}^{)}$ and $b^{(}{'}{}^{,}{''}{}^{)}$ are the frequency-dependent first and second order magneto-phononic coefficients (or magnetic \textit{ionic} Raman tensors). The coupling of coherent infrared-active phonons to the magnetization in first order has been described as phonon inverse Faraday effect, and the coupling in second order as phonon inverse Cotton-Mouton effect \cite{Juraschek2020_3,Juraschek2020}.

In our model, we assume that the terms remaining in the expansion of the projected dynamical matrix are analogous to those in the expansion of the susceptibility for the opto-magnetic effects, setting $a'_{ijk}=b_{ijko}=b''_{ijko}=0$. For the remaining coefficients, $a_{zyx}=-a_{zyx}$ and $b_{zyyz}'=b_{zyzy}'=b_{yzzy}'=b_{yzyz}'$. Using Eqs.~(\ref{eq:phonomagneticHamiltonian}) and (\ref{eq:dynamicalexpansion}), we obtain for the interaction Hamiltonian with coherent infrared-active phonons
\begin{eqnarray}
H_\mathrm{phon} & = & \frac{V_c}{2}\Big[D_{yz}^\mathrm{gs}(\Omega_z)-\mathrm{i} a_{yzx}(\Omega_z) m_x \nonumber \\ 
& + & 2b_{yzzy}'(\Omega_z) l_y l_z\mathcal{Q}_z^*(\Omega_z)e^{i\Omega_z t}+\mathrm{c.c.}\Big]Q_y(t) \nonumber \\
& + & \frac{V_c}{2}\Big[D_{yz}^\mathrm{gs}(\Omega_y)+\mathrm{i} a_{yzx}(\Omega_y) m_x \nonumber \\ 
& + & 2b_{yzzy}'(\Omega_y) l_y l_z\mathcal{Q}_y^*(\Omega_y)e^{i\Omega_y t}+\mathrm{c.c.}\Big]Q_z(t).
\end{eqnarray}
The effective phono-magnetic field acting on the spin $\mathbf{S}_s$ is then given by $\mathbf{B}_s^{\rm eff}=(\gamma \hbar)^{-1}\partial H_{\rm phon}/\partial \mathbf{S}_s$.

To induce an effective phono-magnetic field through the phonon inverse Faraday effect, the coherent phonons have to be elliptically or circularly polarized, for example as a superposition of two phonons with orthogonal polarizations along $z$ and $y$ directions and frequencies $\Omega_z$ and $\Omega_y$, which are excited coherently by an elliptically or circularly polarized laser pulse traveling in $x$ direction. Without loss of generality, the phonon amplitudes can be written as
\begin{eqnarray}
    Q_z(t) & = & \mathcal{Q}_{z}(t)(e^{-{\mathrm i}\Omega_z t}+e^{{\mathrm i}\Omega_z t}), \\
    Q_y(t) & = & \mathcal{Q}_{y}(t)({\mathrm i}e^{-\mathrm{i}\Omega_y t} - {\mathrm i} e^{\mathrm{i}\Omega_y t}),
\end{eqnarray}
where $\mathcal{Q}_{i}(t)$ are the envelopes of the coherently excited phonons that can be obtained by solving the phonon equations of motion. We assume that the projected dynamical matrix is constant during the entire time evolution. The effective phono-magnetic field generated by the phonon inverse Faraday effect, $\mathbf{B}_s^{\rm PIFE}$, through the $a_{yzx}$ term then yields
\begin{eqnarray}\label{eq:PIFEfield}
    \mathbf{B}_s^{\rm PIFE} & = &\frac{V_c}{\gamma \hbar}\mathcal{Q}_{z}(t) \mathcal{Q}_{y}(t) \nonumber\\
    & & \times \left[a_+ \cos(\Omega_-t)+ a_- \cos(\Omega_+t)\right]\hat{x},
\end{eqnarray}
where $\mathbf{B}_1^{\rm PIFE}=\mathbf{B}_2^{\rm PIFE}$, and where we defined $\Omega_\pm = \Omega_z \pm \Omega_y$ and $a_\pm = a_{zyx}(\Omega_y)\pm a_{zyx}(\Omega_z)$. $\mathbf{B}_s^{\rm PIFE}$ can in short be written as cross product and in terms of the angular momentum of the phonons, $\mathbf{Q}\times\dot{\mathbf{Q}}$ \cite{sheng:2006,zhang:2014,Garanin2015,Nakane2018,juraschek2:2017,Juraschek2019,Juraschek2020_3,Juraschek2020,Streib2020}, so that $\mathbf{B}_s^{\rm PIFE}=a_{zyx}\mathbf{Q}\times\mathbf{Q}^*\equiv \tilde{a}_{zyx}\mathbf{Q}\times\dot{\mathbf{Q}}$, where $\mathbf{Q}=(Q_z,Q_y,0)$ and $a_{yzx}(\Omega_i)=\tilde{a}_{yzx}\Omega_i$ \cite{Juraschek2020_3,Juraschek2020}. The prefactors to the difference- and sum-frequency components can therefore be written as $a_\pm = \tilde{a}_{yzx}\Omega_\pm$/2. For circularly polarized phonons (setting $\mathcal{Q}_z=\mathcal{Q}_y\equiv \mathcal{Q}$, $\Omega_z=\Omega_y\equiv \Omega$, and $\tilde{a}_{zyx}\equiv\tilde{a}$) we obtain the regular difference-frequency equation for the effective phono-magnetic field \cite{juraschek2:2017,Juraschek2019,Juraschek2020_3,Juraschek2020},
\begin{equation}\label{eq:PIFEfielddiff}
\mathbf{B}_s^{\rm PIFE} = \frac{V_c}{\gamma\hbar}\mathcal{Q}^2(t)\Omega\tilde{a}\hat{x}.
\end{equation}
The phonon inverse Faraday effect, analogously to the inverse Faraday effect, produces no sum-frequency components in the case of circularly polarized phonons, in which $a_-=\tilde{a}_{yzx}\Omega_-/2=0$. A significant sum-frequency contribution is possible for strongly elliptical phonons with different frequencies for $z$ and $y$ polarizations.


\begin{figure*}[t]
\centering
\includegraphics[scale=0.156]{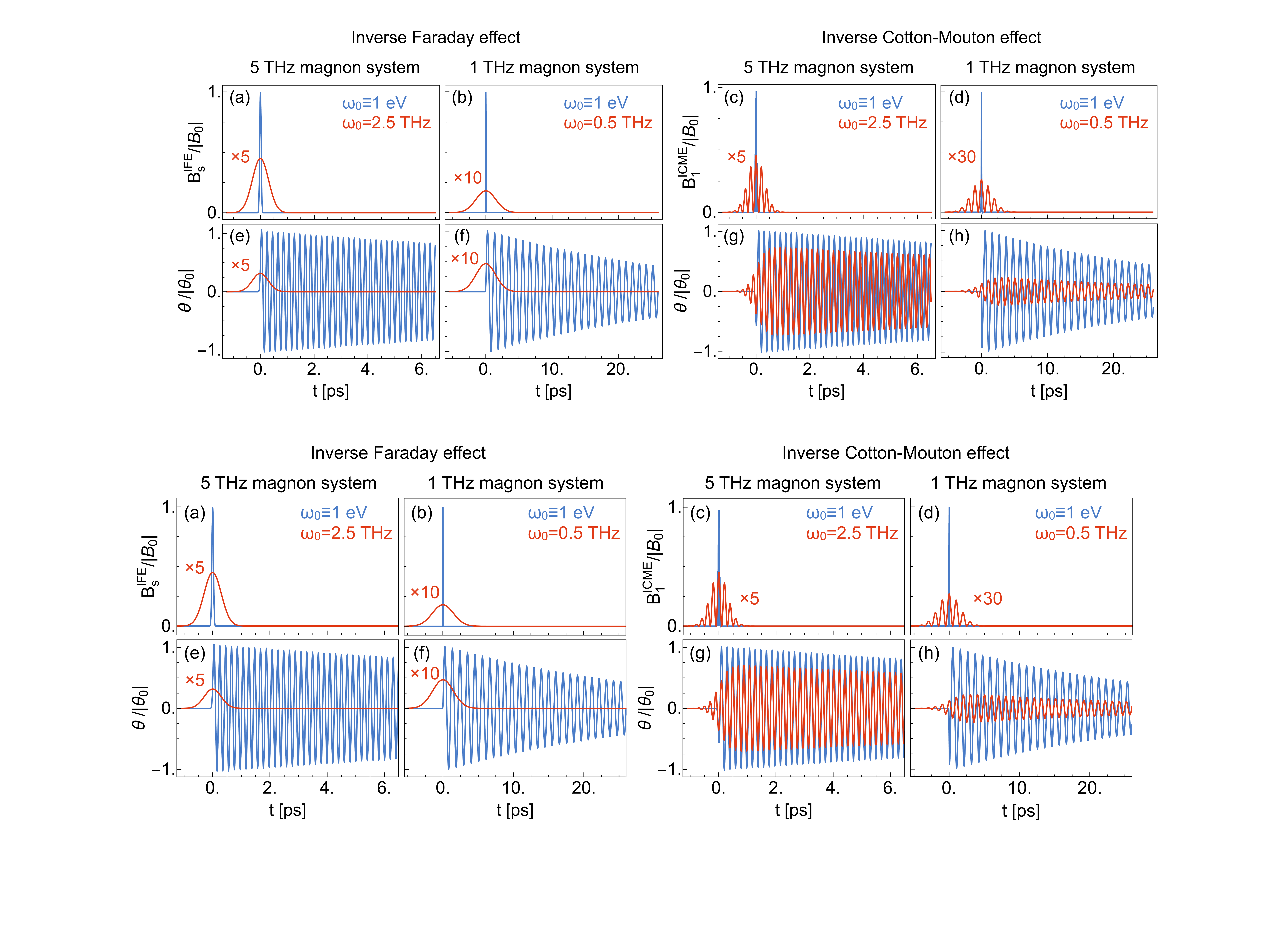}
\caption{
Spin dynamics induced by the inverse Faraday and inverse Cotton-Mouton effects. In (a)-(d), we compare the effective magnetic fields, $B^\mathrm{IFE}_s$ and $B^\mathrm{ICME}_1$, produced by the different optical drives in the near-infrared ($\omega_0\equiv 1$~eV, $\tau= 90$~fs) and terahertz ($\omega_0= 2.5$~THz, $\tau= 1$~ps and $\omega_0\equiv 0.5$~THz, $\tau= 5$~ps) spectral ranges in the 5~THz and 1~THz magnon systems. With these parameters, all three pulses have the same total energy. In (e)-(h) we compare the Faraday rotations, $\theta$, around the $x$-direction arising from the spin dynamics induced by the effective magnetic fields. The Faraday rotations are directly proportional to the induced spin-precession amplitudes, $\theta(t) \propto \mathbf{m}(t)$, see Eq.~(\ref{eq:faradayrotation}). Both the effective magnetic fields and the rotations are normalized to those generated by the respective high-frequency drives, $B_0$ and $\theta_0$.
}
\label{fig:optomagnetism}
\end{figure*}

For an effective phono-magnetic field induced by the phonon inverse Cotton-Mouton effect, we assume that two phonons with orthogonal polarizations $z$ and $y$ are excited in-phase,
\begin{eqnarray}
    Q_z(t) & = & \mathcal{Q}_{z}(t)\cos(\phi)(e^{-{\mathrm i}\Omega_z t}+e^{{\mathrm i}\Omega_z t}),  \\
    Q_y(t) & = & \mathcal{Q}_{y}(t)\sin(\phi)(e^{-{\mathrm i}\Omega_y t}+e^{{\mathrm i}\Omega_y t}),
\end{eqnarray}
where $\phi$ is the angle of linear polarization with respect to the $z$ axis. The effective phono-magnetic fields, $\mathbf{B}_s^{\rm PICME}$, mediated through the $b_{zyyz}'$ term as the only contribution, yield
\begin{eqnarray}\label{eq:PICMEfield}
    \mathbf{B}_{1/2}^{\rm PICME}
    & = & \pm \frac{V_c}{\gamma\hbar} \mathcal{Q}_{z}(t)\mathcal{Q}_{y}(t) b_+ \sin(2\phi)\nonumber \\
    & & \times \left[ \cos(\Omega_+t) + \cos(\Omega_-t) \right] l_z \hat{y},
\end{eqnarray}
where we defined $b_+ = b_{zyyz}'(\Omega_y)+b_{zyyz}'(\Omega_z)$. Analogously to the opto-magnetic case, both difference- and sum-frequency components in the phonon inverse Cotton-Mouton effect scale identically with $b_+$ and the effect is maximized when the polarization of the phonon modes is oriented at a 45 degree angle between the $z$ and $y$ axes.

To evaluate the effective phono-magnetic fields in Eqs.~(\ref{eq:PIFEfield}) and (\ref{eq:PICMEfield}), we calculate the phonon amplitudes $Q_i$ by numerically solving the equations of motion
\begin{equation}\label{eq:phononeom}
\ddot{Q}_i + 2\kappa_i \dot{Q}_i + \Omega_i^2 Q_i = Z_i E_i(t),
\end{equation}
where $\kappa_i$ are the phonon linewidths and $Z_i$ are the mode effective charges \cite{subedi:2014,fechner:2016,Juraschek2018}. The shapes of the electric-field components $E_i$ for the excitation of phonons for the phonon inverse Faraday and phonon inverse Cotton-Mouton effects are given by Eqs.~(\ref{eq:Ez}), (\ref{eq:Ey}), (\ref{eq:Ez_linearlyPolarized}), and (\ref{eq:Ey_linearlyPolarized}), respectively, where the center frequencies are chosen resonant with the phonon frequencies, $\omega_i=\Omega_i$. 

Note that while the coherently excited infrared-active phonons mediate their energy to the spins, they couple nonlinearly to other vibrational degrees of freedom and participate in ionic Raman scattering or two-phonon absorption by other phonons, and may induce transient distortions in the crystal structure that alter the magnetic order quasistatically \cite{juraschek:2017,Radaelli2018,Fechner2018,Gu2018,Khalsa2018,Maehrlein2018,Afanasiev2019,Disa2020,Rodriguez-Vega2020}. These effects and other non-Raman spin-phonon interactions \cite{Fransson2017,Hashimoto2017,Roychoudhury2018,Fechner2018,Streib2018,Nomura2019,Hellsvik2019,Maldonado2019,Berk2019,Streib2019,Ruckriegel2020}, as well as two-magnon processes \cite{Bossini2016,Bossini2019} are not discussed in this work.


\section{Numerical spin-dynamics simulations}

We turn to the numerical evaluations of the spin dynamics that the different optical and phononic drives induce according to the Landau-Lifshitz-Gilbert formalism in Eq.~(\ref{eq:LLG}) and in the pump-probe geometry shown in Fig.~\ref{fig:experiment}. We use two different sets of anisotropy parameters for the antiferromagnetic Heisenberg model described by Eq.~(\ref{eq:groundstate}) in order to create one system with a high-frequency and one system with a low-frequency in-plane magnon mode. For the antiferromagnetic exchange coupling, we set $J=6J_\mathrm{NN}=106$~meV. We set $D_x= 4.3$~meV and $D_y = 1.0$~meV to get an in-plane magnon frequency of 5~THz, and $D_x = 0.43$~meV and $D_y = 0.040$~meV to get an in-plane magnon frequency of 1~THz. In some materials, magnetic anisotropy induces structural distortions along the corresponding lattice directions, which shifts the frequency of the phonons polarized in this direction. As these distortions are generally very small we will neglect the influence of the magnetic anisotropy on the phonon frequencies here.
We set the phenomenological damping of the spin precession to $\Gamma=2.4 \cdot 10^{-4}$ and the opto- and phono-magnetic coefficients to $\alpha_{zyx}=2.8 \times 10^{-58}$, $\beta_{zyyz}'=2.3 \times 10^{-60}$, $a_{zyx} = 4.9 \times 10^{-6}$ m$^3$~s$^{-2}$, $b_{zyyz}'=2.2 \times 10^{-8}$ m$^3$~s$^{-2}$ in the unit system used here. In a real system, the opto- and phono-magnetic coefficients are strongly material dependent and have to be computed using first-principles calculations.

The spin precession induced by the opto- and phono-magnetic fields can be detected through Faraday-rotation experiments, in which the Faraday rotation, $\theta$, is proportional to the oscillation of the ferromagnetic component, $\mathbf{m}$,
\begin{equation} \label{eq:faradayrotation}
    \theta(t) = C \hat{x} \cdot \mathbf{m}(t),
\end{equation}
where the proportionality constant includes the gyromagnetic ratio, $\gamma$, the ion density of the material, and the Verdet constant \cite{Kampfrath2011}. As we compare relative magnitudes of the spin precession, we set $C=1$ without loss of generality.


\subsection{Opto-magnetic effects}

We now compare the relative efficiencies of the difference- and sum-frequency excitation mechanisms via the inverse Faraday and inverse Cotton-Mouton effects for the 5~THz and 1~THz magnon systems. For the laser pulses in Eqs.~(\ref{eq:Ez}), (\ref{eq:Ey}), (\ref{eq:Ez_linearlyPolarized}), and (\ref{eq:Ey_linearlyPolarized}), we assume equal center frequencies and peak electric fields for the $z$ and $y$ components, $\omega_z=\omega_y\equiv\omega_0$ and $\mathcal{E}_{z0}=\mathcal{E}_{y0}\equiv\mathcal{E}_0$. For the difference-frequency components in impulsive stimulated Raman scattering, we use typical pulse parameters in the near-infrared regime, with a photon energy of 1~eV (center frequency $\omega_0=237$~THz), full width at half maximum pulse duration of $\tau=90$ fs, and a peak electric field of $\mathcal{E}_0=25$ MV/cm \cite{Tzschaschel2017}. For the sum-frequency components in two-photon absorption, we require pulses with center frequencies of half the magnon frequency, $\omega_0=\Omega_\mathrm{IP}/2$. We therefore set the parameters to $\omega_0=2.5$~THz, $\tau=1$~ps, and $\mathcal{E}_0=7.5$~MV/cm for the 5~THz magnon system, and $\omega_0=0.5$~THz, $\tau=5$~ps, and $\mathcal{E}_0=3.4$~MV/cm for the 1~THz magnon system, where we chose the pulse durations and peak electric fields such that all three pulses have the same total energy.

We show the normalized effective opto-magnetic fields produced by the inverse Faraday effect according to Eq.~(\ref{eq:IFEfield}) and by the inverse Cotton-Mouton effect according to Eq.~(\ref{eq:ICMEfield}) in Figs.~\ref{fig:optomagnetism}(a)-(d), as well as the normalized Faraday rotations according to Eqs.~(\ref{eq:faradayrotation}) and (\ref{eq:LLG}) in Figs.~\ref{fig:optomagnetism}(e)-(h). In the inverse Faraday effect, only static responses of the effective opto-magnetic fields, $\mathbf{B}^\mathrm{IFE}_s$, resulting from difference-frequency components from both the near-infrared and terahertz pulses show up. While the 1~eV pulse is short enough to impulsivley excite coherent spin precession, the 0.5~THz pulse simply drags the ferromagnetic spin component along for the duration of the pulse. In the inverse Cotton-Mouton effect in contrast, an oscillatory component of the effective opto-magnetic field, $\mathbf{B}^\mathrm{ICME}_1$, is clearly visible. There, the 1~eV as well as both terahertz pulses excite coherent spin precessions with similar magnitudes. The sum-frequency excitations by the terahertz pulses are not impulsive in nature and the spin-precession amplitude builds up gradually, compared to the sudden onset of the amplitude in the difference-frequency case. This behavior of the magnons in the impulsive stimulated Raman scattering and two-photon absorption processes underlying the inverse Cotton-Mouton effect is equivalent to that of Raman-active phonons \cite{Juraschek2018}.


\begin{figure*}[t]
\centering
\includegraphics[scale=0.156]{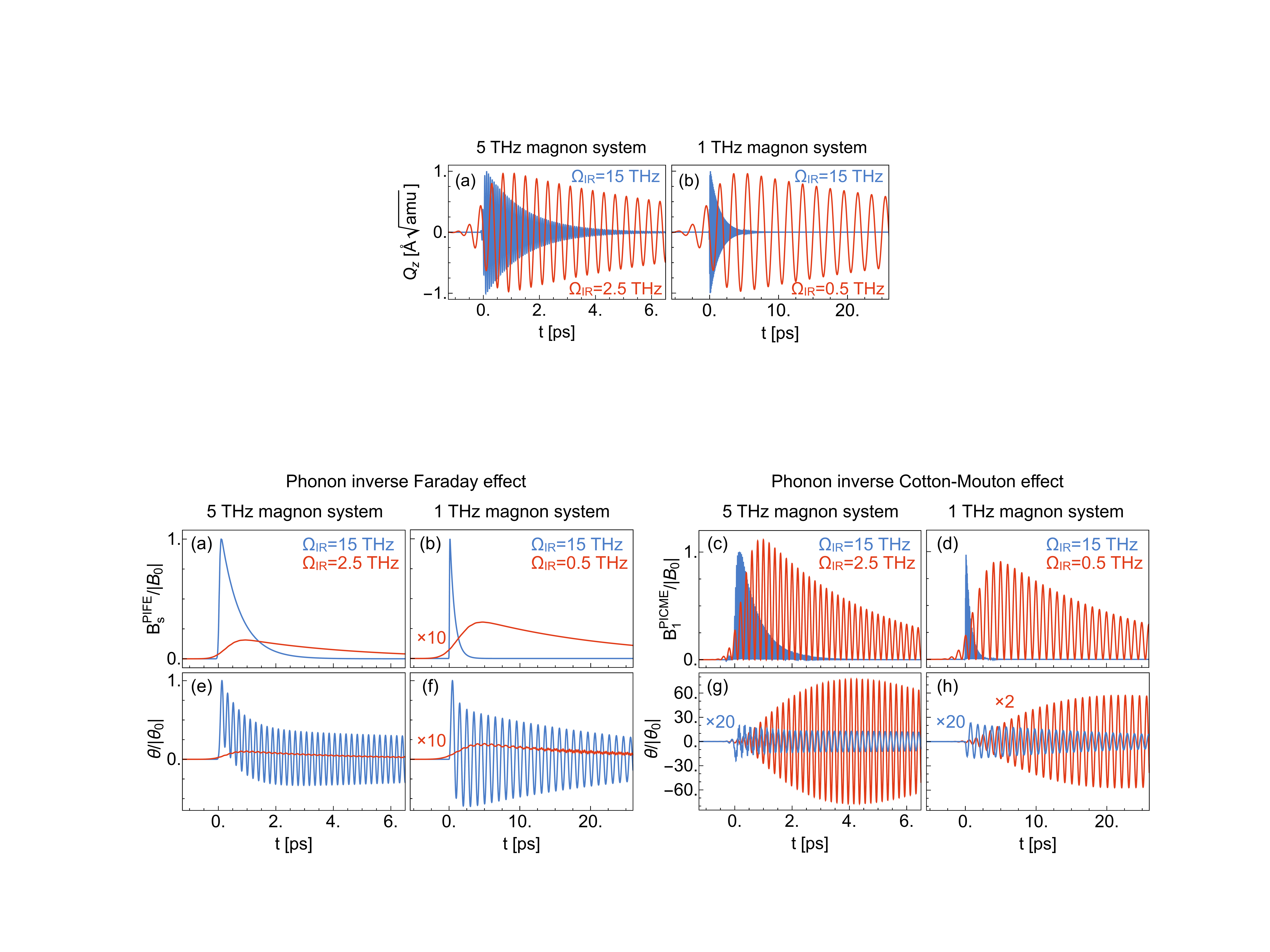}
\caption{
Spin dynamics induced by the phonon inverse Faraday and phonon inverse Cotton-Mouton effects. In (a)-(d), we compare the effective magnetic fields, $B^\mathrm{PIFE}_s$ and $B^\mathrm{PICME}_1$, produced by the different phonons (15, 2.5, and 0.5~THz) in the 5~THz and 1~THz magnon systems. In (e)-(h) we compare the Faraday rotations, $\theta$, around the $x$-direction arising from the spin dynamics induced by the effective phono-magnetic fields. The Faraday rotations are directly proportional to the induced spin-precession amplitudes, $\theta(t) \propto \mathbf{m}(t)$, see Eq.~(\ref{eq:faradayrotation}). Both the effective magnetic fields and the rotations are normalized to those generated by the respective high-frequency drives, $B_0$ and $\theta_0$.
}
\label{fig:phonomagnetism}
\end{figure*}


\subsection{Phono-magnetic effects}

We next compare the relative efficiencies of the difference- and sum-frequency excitation mechanisms via the phonon inverse Faraday and phonon inverse Cotton-Mouton effects for the 5~THz and 1~THz magnon systems. In a real material, the coherent excitation of infrared-active phonons imposes an additional limitation on the possible excitation strength of the laser pulses. When the amplitude of atomic motion exceeds a certain threshold, the structure of the crystal will break down. 
It is therefore more meaningful to compare the phono-magnetic effects in terms of equal phonon amplitudes and not in terms of equal total pulse energies as in the case of the opto-magnetic effects. 


\begin{figure}[b]
\centering
\includegraphics[scale=0.15]{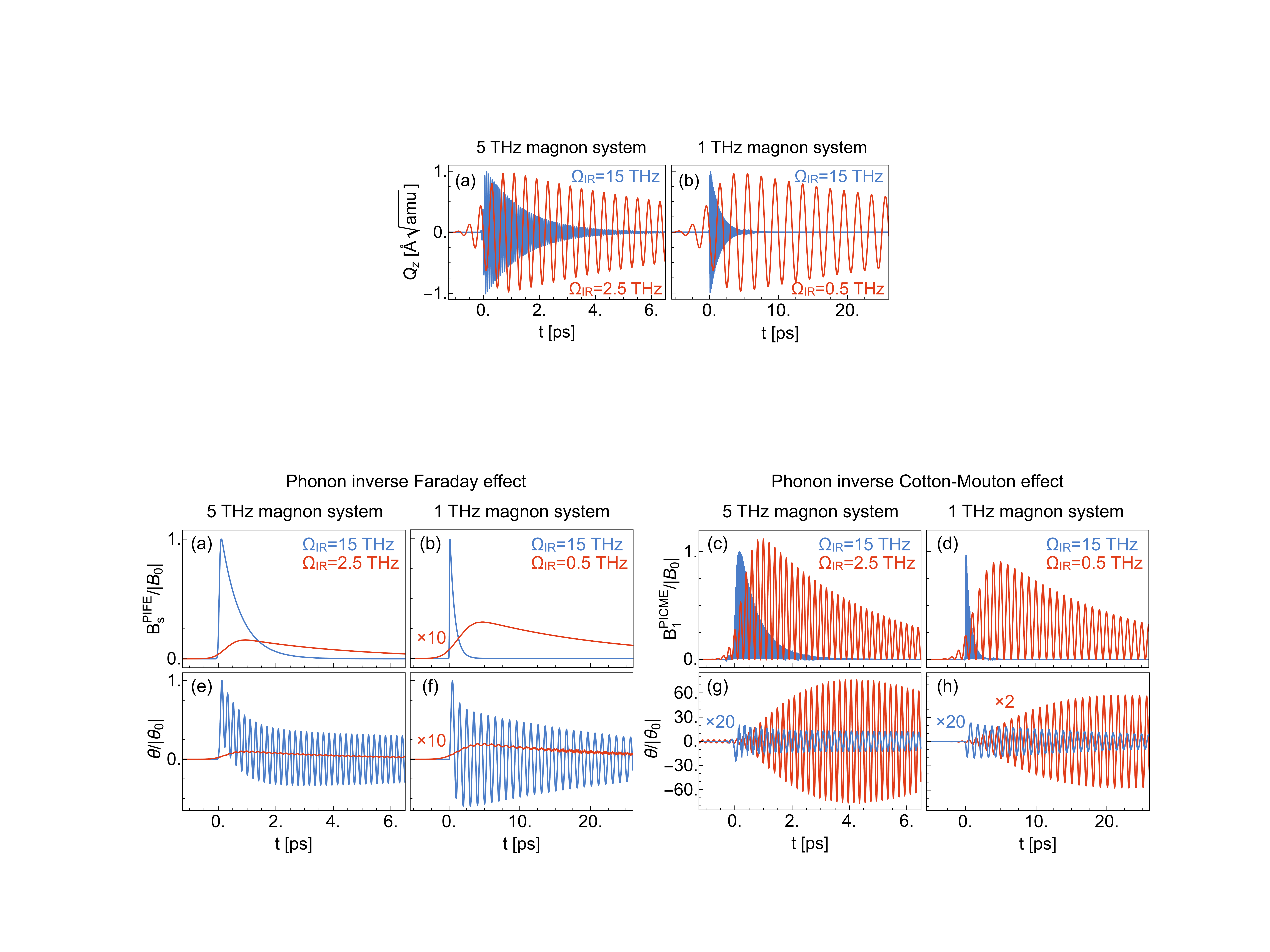}
\caption{
Time evolution of the phonon amplitudes, $Q_z$, in response to the excitations by resonant terahertz pulses. (a) 15 and 2.5~THz phonons in the 5~THz magnon system, and (b) 15 and 0.5~THz phonons in the 1~THz magnon system. We chose the pulse energies such that the maxima of $Q_z$ are equal for the high- and low-frequency phonons, respectively. In the case of circular excitation, $Q_y$ (not shown) is shifted in phase by $\pi/4$ with respect to $Q_z$ and in the case of linear excitation it is in phase with $Q_z$.
}
\label{fig:phonons}
\end{figure}

For the difference-frequency components in ionic Raman scattering, we assume phonons polarized along the $z$ and $y$ directions with equal eigenfrequencies of $\Omega_z=\Omega_y\equiv\Omega_\mathrm{IR}=15$~THz, which are resonantly driven by a pulse with a center frequency of $\omega_0=\Omega_\mathrm{IR}$, a full width at half maximum pulse duration of $\tau=0.1$~ps, and a peak electric field of $\mathcal{E}_0=23.7$~MV/cm. For the sum-frequency components in two-phonon absorption, the phonon frequencies have to be half the magnon frequency, $\Omega_\mathrm{IR}=\Omega_\mathrm{IP}/2$, at 2.5~THz for the 5~THz magnon system and at 0.5~THz for the 1~THz magnon system. We yield phonon amplitudes equal to that of the 15~THz phonon when we set the parameters as follows: $\omega_0=\Omega_\mathrm{IR}=2.5$~THz, $\tau=1$~ps, and $\mathcal{E}_0=0.4$~MV/cm for the 5~THz magnon system, and $\omega_0=\Omega_\mathrm{IR}=0.5$~THz, $\tau=5$~ps, and $\mathcal{E}_0=16$~kV/cm for the 1~THz magnon system. The excitation of the 2.5~THz phonon with equal amplitude as the 15~THz phonon therefore requires a factor of 350 less total pulse energy $\propto \mathcal{E}_0^2 \tau$, and even a factor of 40'000 less for the 0.5~THz phonon. For the linewidths of all phonons, we use phenomenological values of $\kappa_z=\kappa_y\equiv\kappa_\mathrm{IR}=0.05\times\Omega_\mathrm{IR}$. We further set the mode effective charges to $Z_\mathrm{IR}=1$~$e\ang{}$, where $e$ is the elementary charge \cite{Juraschek2018}.

We show the amplitudes of the coherently excited infrared-active phonons, $Q_z$, according to Eq.~(\ref{eq:phononeom}) in Figs.~\ref{fig:phonons}(a) and (b). In the case of circular excitation, $Q_y$ (not shown in the figures) is shifted in phase by $\pi/4$ with respect to $Q_z$ and in the case of linear excitation it is in phase with $Q_z$. We further show the normalized effective phono-magnetic fields produced by the phonon inverse Faraday effect according to Eq.~(\ref{eq:PIFEfield}) and by the phonon inverse Cotton-Mouton effect according to Eq.~(\ref{eq:PICMEfield}) in Figs.~\ref{fig:phonomagnetism}(a)-(d), as well as the normalized Faraday rotations according to Eqs.~(\ref{eq:faradayrotation}) and (\ref{eq:LLG}) in Figs.~\ref{fig:phonomagnetism}(e)-(h). 

The phonon inverse Faraday effect displays only static responses of the effective phono-magnetic fields, $\mathbf{B}^\mathrm{PIFE}_s$, resulting from difference-frequency components from the 15, 2.5, and 0.5~THz phonons, analogously to its opto-magnetic counterpart. Analogously to the opto-magnetic case, the high-frequency drive of the 15~THz phonon leads to an impulsive excitation of the magnon through ionic Raman scattering, while the 2.5 and 0.5~THz phonons induce a transient magnetization instead, as the ferromagnetic component $\mathbf{m}$ follows the envelope of the respective phonon. Notably, also the 15~THz phonon induces a transient magnetization that is reminiscent of the transient structural distortion generated through nonlinear phonon-phonon coupling in nonlinear phononics experiments \cite{forst:2011,Forst2013,subedi:2014,Mankowsky:2015}. 

For the phonon inverse Cotton-Mouton effect, a strong oscillatory component is visible in the effective phono-magnetic field, $\mathbf{B}^\mathrm{PICME}_1$. While all three phonons are excited with similar amplitudes and yield similar magnitudes of effective magnetic fields, the spin-precession amplitudes and therefore Faraday rotations induced by the 2.5 and 0.5~THz phonons exceed those of the 15~THz phonon by a factor of 30-50. This is due to the non-impulsive nature of the sum-frequency process that gradually builds up the amplitude of the magnon, and which benefits from the long lifetime of the low-frequency phonons, while the impulsive excitation through the difference-frequency components is independent of the envelope of the high-frequency phonon. Analog to the opto-magnetic case, the behavior of the magnons in the ionic Raman scattering and two-phonon absorption processes underlying the phonon inverse Cotton-Mouton effect is similar to that of Raman-active phonons \cite{Juraschek2018}.


\section{Discussion}

Our results show that the inverse Faraday and phonon inverse Faraday effects do not produce sum-frequency components for circularly polarized light or phonons, respectively. The weak dependence of the first-order opto-magnetic coefficients on the frequency of light and phonons makes it further unlikely that significant excitation will be achieved through elliptical polarization. Intriguingly however, the difference-frequency components of the phonon inverse Faraday effect induce a transient magnetization that could be considered a magnetic analogue of the transient structural distortion in nonlinear phononics \cite{forst:2011,Forst2013,subedi:2014,Mankowsky:2015,subedi:2015,fechner:2016,juraschek:2017,Radaelli2018,Fechner2018,Gu2018,Khalsa2018,Maehrlein2018,Disa2020,Rodriguez-Vega2020}. 

The inverse Cotton-Mouton and phonon inverse Cotton-Mouton effects in contrast produce strong sum-frequency components that scale identically in terms of the second-order opto- and phono-magnetic coefficients as the difference-frequency components. Here, for the same total pulse energies and magnitudes of the opto-magnetic coefficients, similar spin-precession amplitudes can be achieved through impulsive stimulated Raman scattering and two-photon absorption. In the phono-magnetic effects, spin-precession amplitudes induced by two-phonon absorption are at least an order of magnitude larger than those induced by ionic Raman scattering for equally strong phonon excitations and phono-magnetic coefficients. This difference arises from the non-impulsive nature of the sum-frequency mechanisms in combination with the long lifetimes of the low-frequency phonons, where the spin-precession amplitude can build up gradually over time. In addition, the pulse energies required to excite these low-frequency phonons are orders of magnitude lower than the ones required to excite high-frequency phonons, making the phonon inverse Cotton-Mouton effect particularly efficient.

From a feasibility perspective, opto-magnetic sum-frequency excitation can straightforwardly be implemented by choosing the center frequency of the laser pulse at half of the magnon frequency. The feasibility of the phono-magnetic sum-frequency mechanism in turn strongly depends on the material properties that determine the magnon and phonon frequencies. For low-frequency magnons, as the example of the 1~THz magnon here shows, phonons with an eigenfrequency of 0.5~THz would be required that are not present in most existing compounds. The phono-magnetic sum-frequency mechanism should therefore be more applicable to high-frequency magnons, for example in complex magnetic structures such as yttrium iron garnet \cite{Princep2017}, which host a large number of magnon and phonon modes. There, the high selectivity provided by the sum-frequency mechanisms due to their resonant conditions in combination with the high selectivity of coherent phonon excitations achievable in recent years \cite{liu2020} may become powerful tools for spin dynamical control in the future.


\begin{acknowledgements}
We acknowledge fruitful discussions with Sebastian Maehrlein, Alexandra Kalashnikova, Tobias Kampfrath, Christian Tzschaschel and Tom\'{a}\v{s} Neuman. D.M.J. and D.S.W. contributed equally to this work. This work was supported by the Swiss National Science Foundation under Project No. 184259, the DARPA DRINQS program under Award No. D18AC00014, as well as by the Department of Energy ‘Photonics at Thermodynamic Limits Energy Frontier Research Center under grant number DE-SC0019140. D.S.W. is supported by a NSF Graduate Research Fellowship. P.N. is a Moore Inventor Fellow through Grant GBMF8048 from the Gordon and Betty Moore Foundation.
\end{acknowledgements}



%

\end{document}